\documentclass[aps,prd,superscriptaddress,preprint,nopreprintnumbers,nofootinbib,showpacs]{revtex4}
\usepackage{graphicx,epsfig}
\usepackage{dcolumn}
\usepackage{bm}

\begin{document}

\title{\boldmath Branching fraction measurements of
$\chi_{c0}$ and $\chi_{c2}$ to $\pi^0\pi^0$
and $\eta\eta$}

\author{
M.~Ablikim$^{1}$, M.~N.~Achasov$^{5}$, L.~An$^{9}$, Q.~An$^{31}$, Z.~H.~An$^{1}$, J.~Z.~Bai$^{1}$, Y.~Ban$^{18}$, N.~Berger$^{1}$, J.~M.~Bian$^{1}$, I.~Boyko$^{13}$
% ...
\\
\vspace{0.2cm}
(BESIII Collaboration)\\
{\it
$^{1}$ Institute of High Energy Physics, Beijing 100049, P. R. China\\
% ...
$^{5}$ G.I. Budker Institute of Nuclear Physics SB RAS (BINP), Novosibirsk 630090, Russia\\
% ...
$^{9}$ Guangxi University, Naning 530004, P. R. China\\
% ...
$^{13}$ Joint Institute for Nuclear Research, 141980 Dubna, Russia\\
% ...
$^{18}$ Peking University, Beijing 100871, P. R. China\\
% ...
$^{31}$ University of Science and Technology of China, Hefei 230026, P. R. China\\
% ...
} } \noaffiliation
\date{\today}

\begin{abstract}
Using a sample of $1.06 \times 10^8$ $\psi^{\prime}$ decays
collected by the BESIII detector, $\chi_{c0}$ and $\chi_{c2}$ decays
into $\pi^0\pi^0$ and $\eta\eta$ are studied. The branching fraction
results are ~$Br(\chi_{c0}\to\pi^0\pi^0)=(3.23\pm 0.03\pm0.23 \pm
0.14)\times 10^{-3}$,~$Br(\chi_{c2}\to\pi^0\pi^0)=(8.8\pm 0.2\pm
0.6\pm0.4 )\times 10^{-4}$,~$Br(\chi_{c0}\to\eta\eta)=(3.44\pm
0.10\pm0.24 \pm0.2 )\times 10^{-3}$, and
$Br(\chi_{c2}\to\eta\eta)=(6.5\pm 0.4\pm 0.5\pm 0.3)\times 10^{-4}$,
where the uncertainties are statistical, systematic due to this
measurement, and systematic due to the branching fractions of
$\psi^{\prime}\to\gamma\chi_{cJ}$, respectively. The results provide
information on the decay mechanism of $\chi_c$ states into
pseudoscalars.

\end{abstract}


\maketitle
\section{\boldmath Introduction}

In the quark model, the $\chi_{cJ}$ ($J=0,1,2$) mesons are $L=1$
$c\bar{c}$ states. Since they cannot be produced directly in
$e^+e^-$ collisions, they are not as well studied as the $\psi$
states. On the other hand, $\psi^{\prime}\to\gamma\chi_{cJ}$ decays
yield many $\chi_{cJ}$ mesons, providing a clean environment for
$\chi_{cJ}$ investigations. In this paper, we study two-body decays
of the $\chi_{c0}$ and $\chi_{c2}$ into $\pi^0\pi^0$ and $\eta\eta$
final states\footnote{We do not consider $\chi_{c1}$ decays into
these final states, as they are forbidden by spin-parity
conservation.}. Knowledge gained from these decays provides
information on both the $\chi_{cJ}$ parents and their pseudo-scalar
daughters, as well as a greater understanding of the decay
mechanisms of $\chi_{cJ}$ mesons \cite{ZHOU}.

Recently, $\chi_{c0}$ and $\chi_{c2}$ decays into two-meson final
states were studied by the CLEOc collaboration \cite{bib:cleoc}. In
this analysis, we use a sample of $1.06 \times 10^8$ $\psi^{\prime}$
decays collected by the BESIII detector to perform a study of these
decays.

\section {\boldmath BESIII and BEPCII}
 The analysis reported here is based on about
$1.06 \times 10^8 ~\psi^{\prime}$ events collected by BESIII at
BEPCII. BEPCII/BESIII~\cite{bes3} is a major upgrade of the BESII
experiment at the BEPC accelerator~\cite{bes2} for studies of hadron
spectroscopy and $\tau$-charm physics \cite{bes3phys}. The design
peak luminosity of the double-ring $e^+e^-$ collider, BEPCII, is
$10^{33}$ cm$^{-2}s^{-1}$ at a beam current of 0.93~A.  The BESIII
detector with a geometrical acceptance of 93\% of 4$\pi$, consists
of the following main components: 1) a small-celled, helium-based
main draft chamber (MDC) with 43 layers.
The average single wire resolution is 135~$\mu m$, and the momentum
resolution for 1~GeV/$c$ charged particles in a 1~T magnetic field is
0.5\%; 2) an electromagnetic calorimeter (EMC) made of 6240 CsI (Tl)
crystals arranged in a cylindrical shape (barrel) plus two endcaps.
For 1.0~GeV photons, the energy resolution is 2.5\% in the barrel
and 5\% in the endcaps, and the position resolution is 6~mm in the
barrel and 9~mm in the endcaps; 3) a Time-Of-Flight system (TOF) for
particle identification composed of a barrel part made of two layers
with 88 pieces of 5~cm thick, 2.4~m long plastic scintillators in
each layer, and two endcaps with 96 fan-shaped, 5~cm thick, plastic
scintillators in each endcap.  The time resolution is 80~ps in the
barrel, and 110~ps in the endcaps,  corresponding to better than a 2 sigma K/pi separation for momenta below about 1~GeV/$c$; 4) a muon
chamber system (MUC) made of 1000~m$^2$ of Resistive Plate Chambers
(RPC) arranged in 9 layers in the barrel and 8 layers in the endcaps
and incorporated in the return iron of the superconducting magnet.
The position resolution is about 2~cm.


 The optimization of the event selection and the estimation of physics
 backgrounds are performed through Monte Carlo simulations. The
 GEANT4-based simulation software BOOST~\cite{boost} includes the
 geometric and material description of the BESIII detectors, the
 detector response and digitization models, as well as the tracking of
 the detector running conditions and performance.  The production
 of the $\psi^{\prime}$ resonance is simulated by the Monte Carlo event
 generator KKMC~\cite{kkmc}, while the decays are generated by
 EvtGen~\cite{besevtgen} for known decay modes with branching ratios
 being set to the PDG~\cite{PDG} world average values,
 and by Lundcharm~\cite{lund} for the remaining unknown decays.  The
 analysis is performed in the framework of the BESIII Offline Software
 System~(BOSS)~\cite{boss} which takes care of the detector
 calibration, event reconstruction and data storage.

\section {\boldmath Event selection}
\label{event_selection}
A photon candidate is defined as a shower in the EMC with an energy
deposit exceeding 50~MeV.
The $\pi^0$ and $\eta$ candidates are reconstructed from pairs of
photon candidates, using the average event vertex of each run as the
assumed origin of the photons. For $\pi^0\to\gamma\gamma$, the
$\gamma\gamma$ invariant mass is required to satisfy
$0.075$~GeV/$c^2<M(\gamma\gamma)<0.175$~GeV/$c^2$. For
$\eta\to\gamma\gamma$, the $\gamma\gamma$ invariant mass is required
to satisfy $0.458$~GeV/$c^2<M(\gamma\gamma)<0.608$~GeV/$c^2$. The
decay angle of a photon is the polar angle measured in the $\pi^0$
or $\eta$ rest frame with respect to the $\pi^0$ or $\eta$ direction
in the $\psi^{\prime}$ rest frame. Real $\pi^0$ and $\eta$ mesons
decay isotropically, and their angular distributions are flat.
However, the $\pi^0$ and $\eta$ candidates that originate from a
wrong photon combination do not have a flat distribution in this
variable. To remove wrong photon combinations, the decay angle is
required to satisfy $|\cos\theta_{decay}|<0.95$.

Candidate events for the final states of interest ($\gamma
\pi^0\pi^0$ and $\gamma \eta\eta$ ) are selected using the following
basic selection criteria. An event must have 5 or 6 photons and no
charged tracks. All possible two photon pairings (the radiative
photon from the $\psi^{\prime}$ decay which has $E<0.4$~GeV is
excluded) in the event are used to form $\pi^0$ and $\eta$
candidates. The candidate event uses the photon pairings giving the
minimum

\[ \chi_{\pi^0\pi^0/\eta\eta}=\sqrt{P_1^2(\pi^0/\eta) +  P_2^2(\pi^0/\eta)} , \]

\noindent with $P_1$ and $P_2$ being the pulls, defined as:

\[ P(\pi^0/\eta) = \left [ M_{\gamma\gamma} - m_{\pi^0/\eta} \right ] /
                         \sigma_{\gamma\gamma}, \]

\noindent where $M_{\gamma\gamma}$ is the reconstructed
$\gamma\gamma$ invariant mass, $m_{\pi^0/\eta}$ is the known $\pi^0$
or $\eta$ mass~\cite{PDG}, and $\sigma_{\gamma\gamma}$ is the
$\gamma\gamma$ mass resolution, with typical values of 7~MeV/$c^2$
for the $\pi^0$ and 12~MeV/$c^2$ for the $\eta$. If there is more
than one radiative photon candidate (E$<$0.4~GeV), the one that
gives the least $|M_{5\gamma}-m_{\psi^{\prime}}|$ is used.

Backgrounds with missing final state particles are suppressed
by requiring small transverse momentum squared $p_{t\gamma}^2$,

\[ p_{t\gamma}^2=4p_{miss}^2\sin^2(\theta_{\gamma}/2), \]

\noindent where $p_{miss}$ is the missing momentum opposite to the
$\pi^0\pi^0$ or $\eta\eta$ system and $\theta_{\gamma}$ is the angle
between the radiative photon and the direction of the missing
momentum $p_{miss}$. The $\gamma\pi^0\pi^0$ events are required to
satisfy $ p_{t\gamma}^2<0.04$~(GeV/$c)^2$, while the $\gamma\eta\eta$
events are required to satisfy $ p_{t\gamma}^2<0.01$~(GeV/$c)^2$ and
$\chi_{\eta\eta}<4$.

To study the efficiency of the $\psi^{\prime}\to\gamma\chi_{cJ},
\chi_{cJ}\to\pi^0\pi^0$ and $\chi_{cJ}\to\eta\eta$ selection, Monte
Carlo samples for each $\chi_{cJ}$ state into each final state are
generated using a $(1+\lambda \cos^2 \theta)$ distribution, where
$\theta$ is the radiative photon angle relative to the positron beam
direction, and $\lambda=1$  for $\chi_{c0}$ and $\lambda= 1/13$ for
$\chi_{c2}$, in accordance with expectations for E1 transitions. The
decay products of the $\chi_{c0}$ are generated using a flat angular
distribution, while those of the $\chi_{c2}$ are generated according
to a double correlation function of the polar angles of the mesons
measured in the $\chi_c$ rest frame relative to the transition
photon direction ~\cite{besevtgen,DC}. The efficiencies obtained
from the Monte Carlo simulation are shown in Table~\ref{eff}.
\begin{table}[htb]
\caption{Efficiencies (in \%) obtained from analysis of Monte Carlo
generated events.}

\begin{tabular}{|l|c|c|c|c|c|c|c|}
\hline
Mode  &  $\chi_{c0}$  &  $\chi_{c2}$ \\

 \hline
$\pi^0\pi^0$     & $55.6 \pm0.2$& $59.8\pm 0.2$    \\
$\eta \eta$      & $40.3 \pm0.2$ & $43.9\pm 0.2$    \\

\hline
\end{tabular}
   \label{eff}
\end{table}

\begin{figure}[htb]
   \centerline{
   \psfig{file=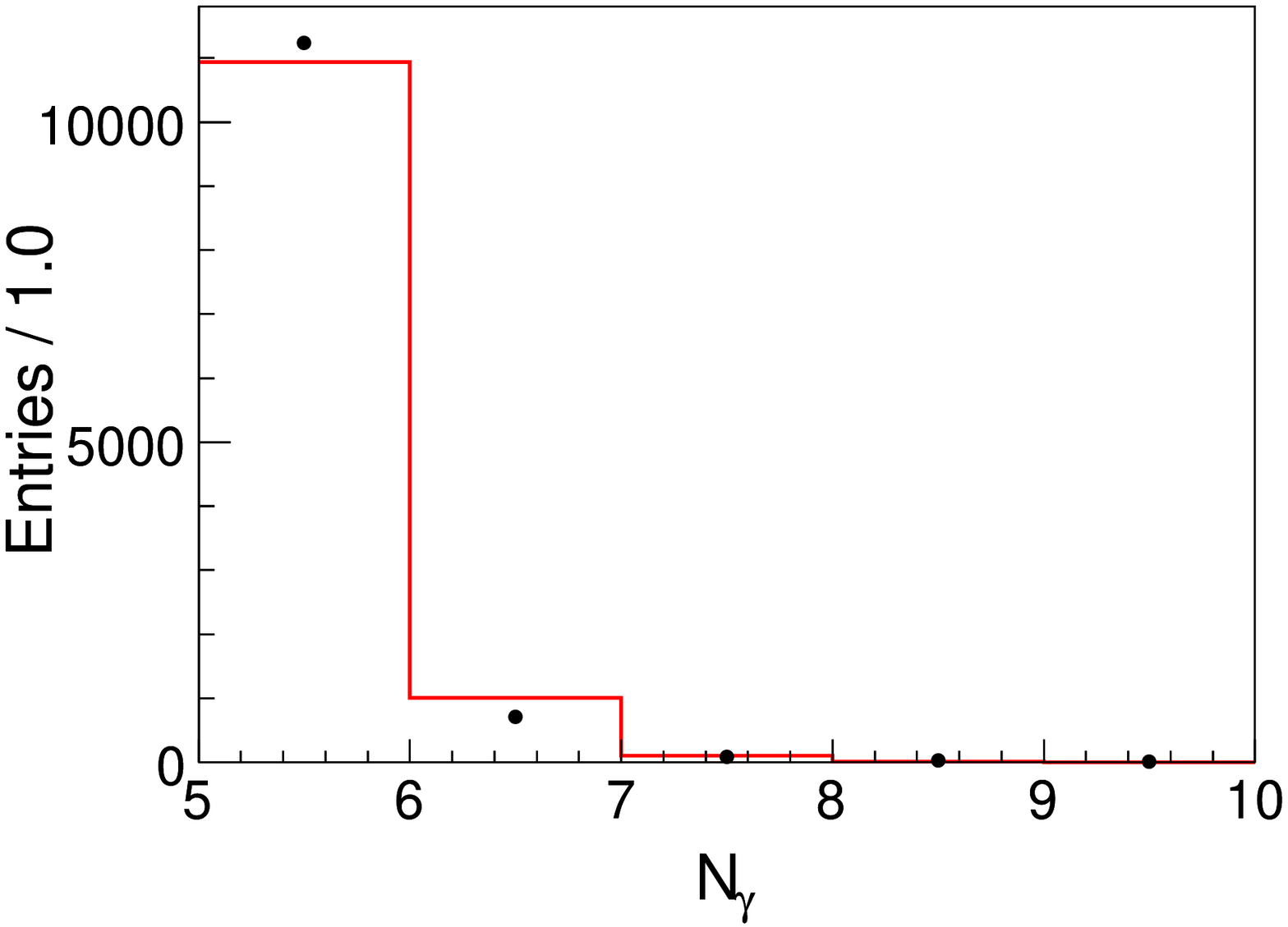,width=8cm,angle=0}
              \put(-30,120){(a)}
   \psfig{file=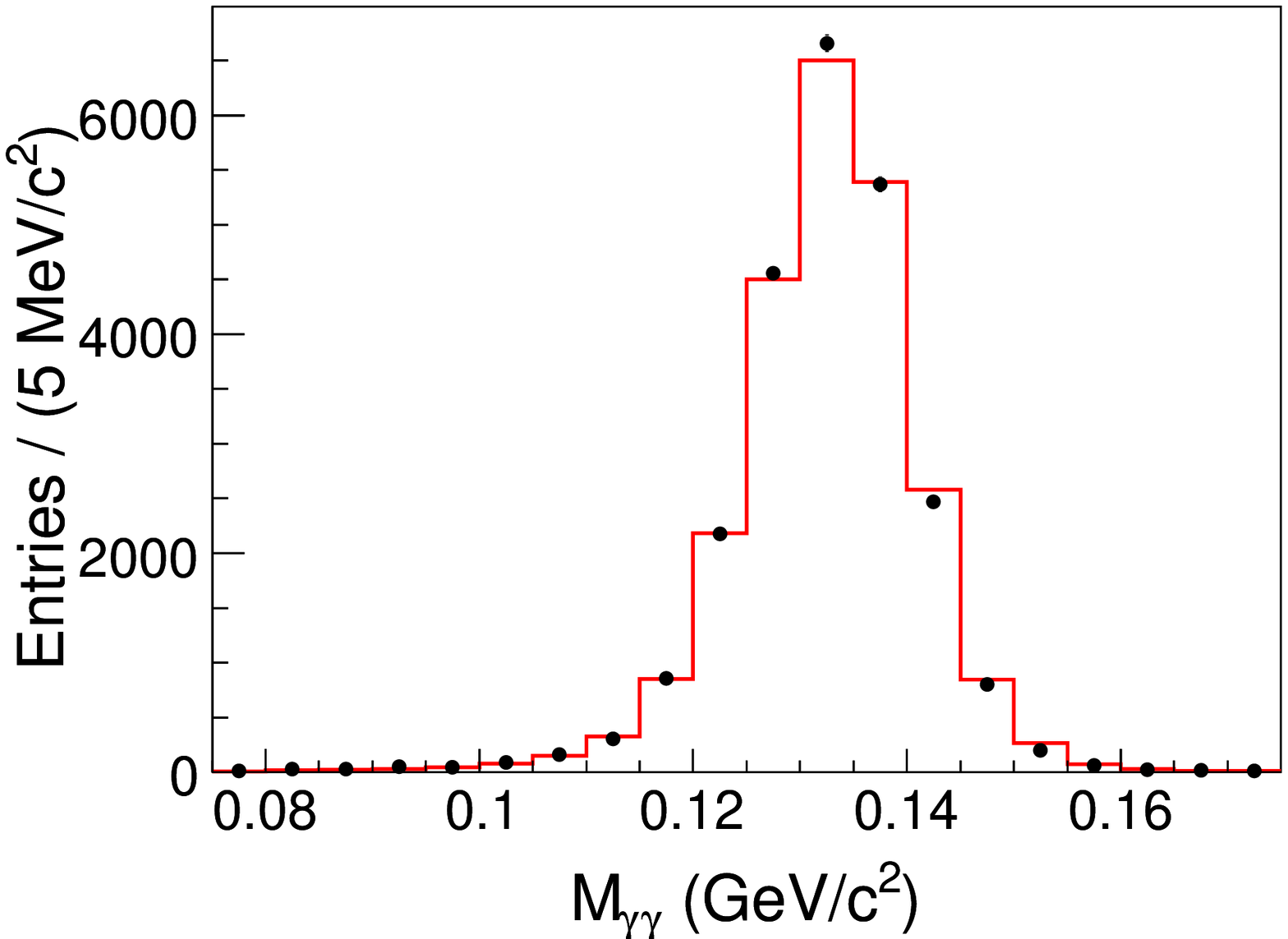,width=8cm,angle=0}
              \put(-30,120){(b)}}
                \centerline{
   \psfig{file=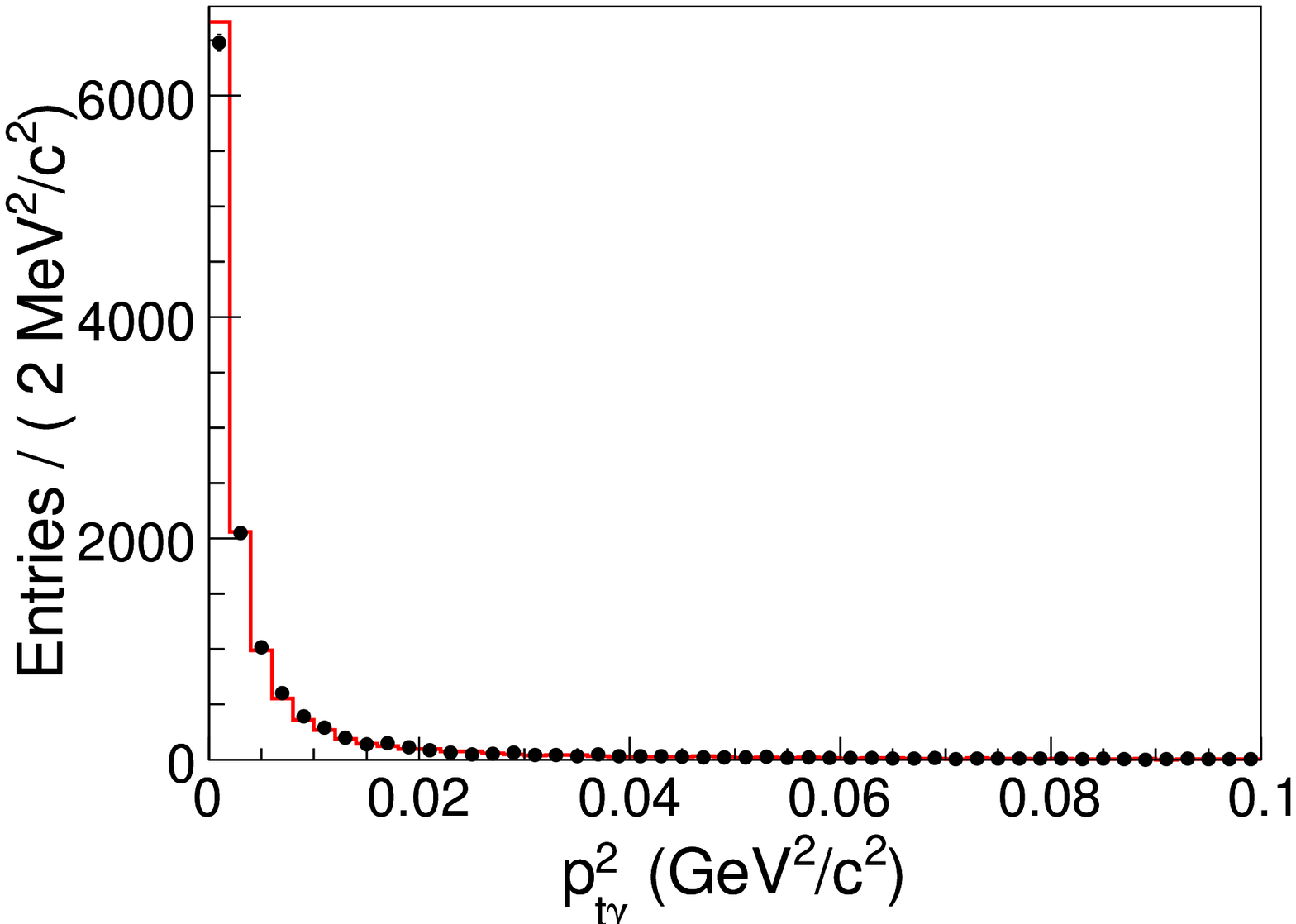,width=8cm,angle=0}
              \put(-30,120){(c)}
   \psfig{file=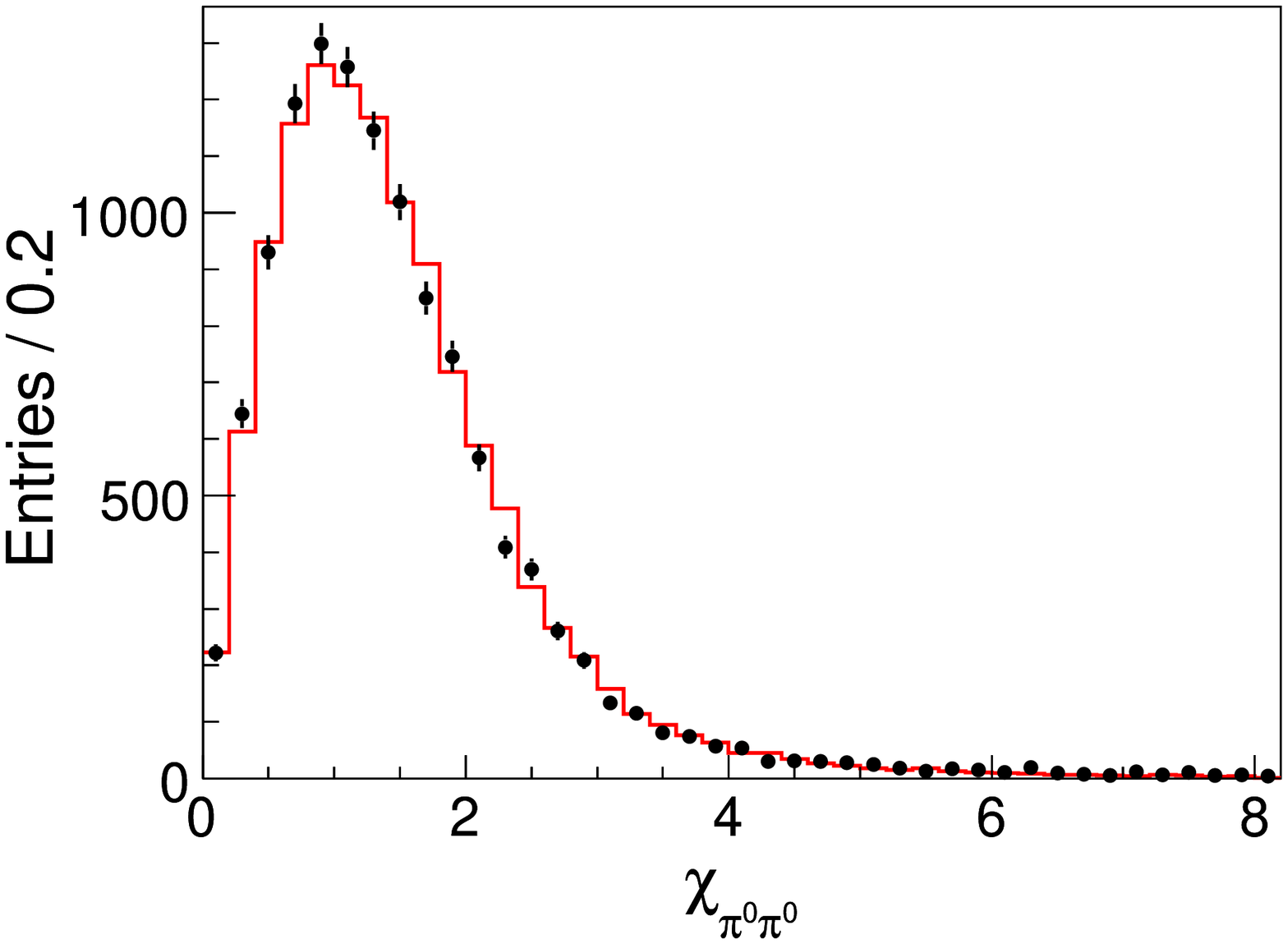,width=8cm,angle=0}
              \put(-30,120){(d)}}
   \caption{Comparisons between data and Monte Carlo simulation of $\psi^{\prime}\to\gamma\chi_{cJ},
\chi_{cJ}\to\pi^0\pi^0$ for selection criteria used.
   (a) Photon multiplicity distribution.
   (b) The $\gamma\gamma$ invariant mass distribution for $\pi^0$
candidates.
   (c) The distribution of $p_{t\gamma}^2$.
   (d) The $\chi_{\pi^0\pi^0}$ distribution.
   Dots with error bars are data in the $\chi_{c0}$
region. The histogram is the Monte Carlo simulation for
$\psi^{\prime} \to \gamma\chi_{c0}, \chi_{c0}\to\pi^0\pi^0$ plus
normalized background estimated from inclusive $\psi^{\prime}$ Monte
Carlo samples.}
   \label{comp_pp}
\end{figure}

\begin{figure}[htb]
   \centerline{
   \psfig{file=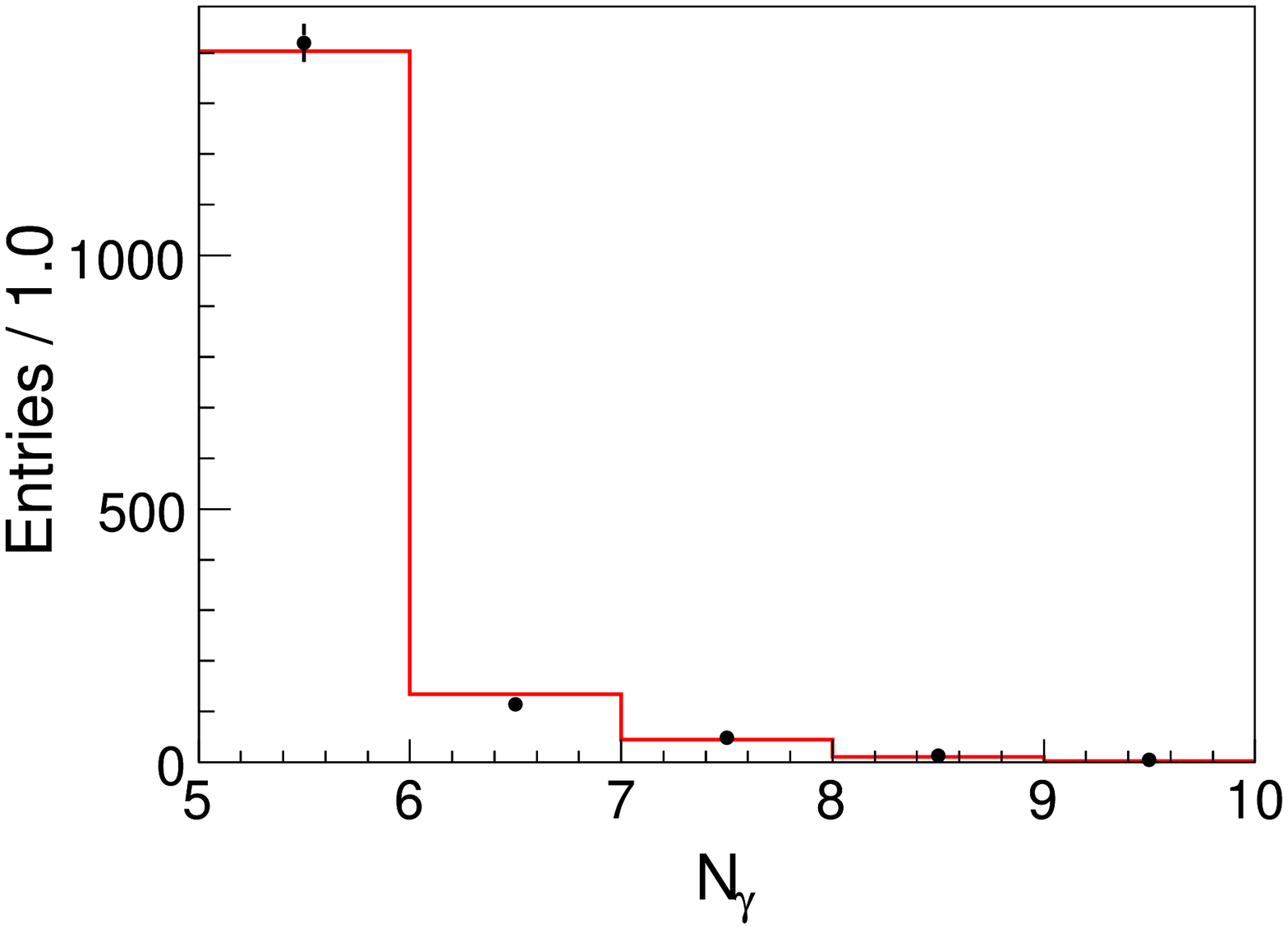,width=8cm,angle=0}
              \put(-30,120){(a)}
   \psfig{file=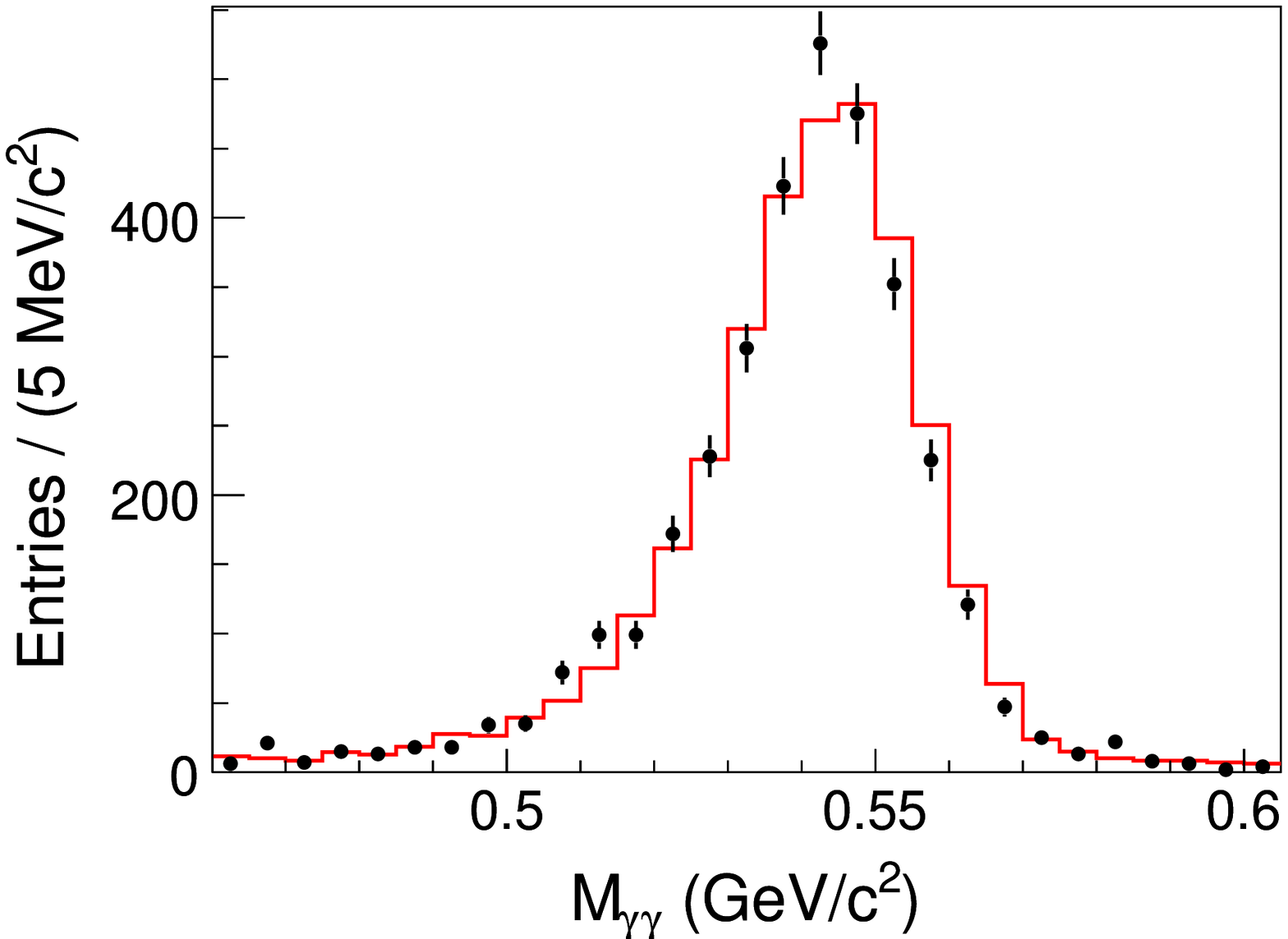,width=8cm,angle=0}
              \put(-30,120){(b)}}
                \centerline{
   \psfig{file=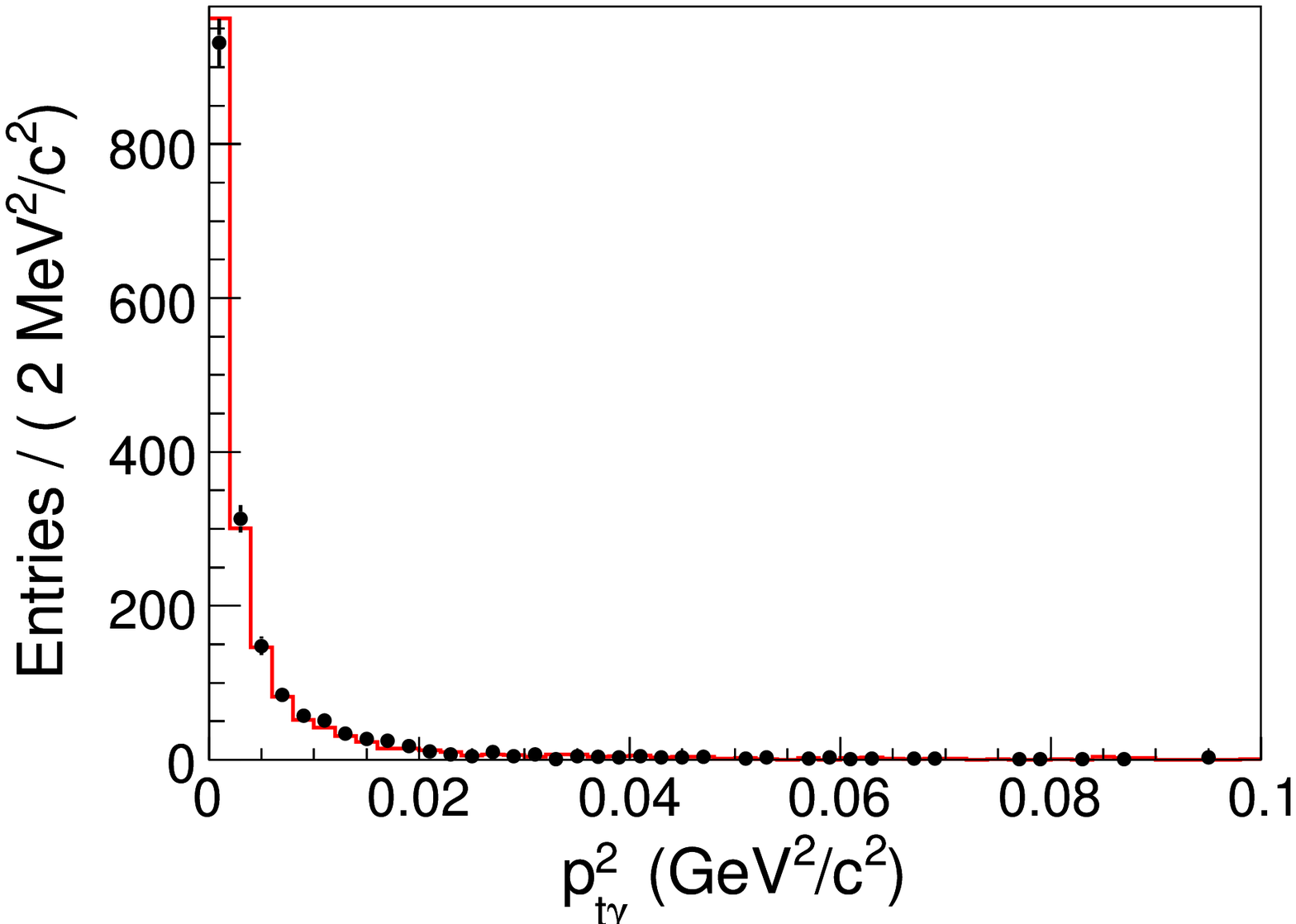,width=8cm,angle=0}
              \put(-30,120){(c)}
   \psfig{file=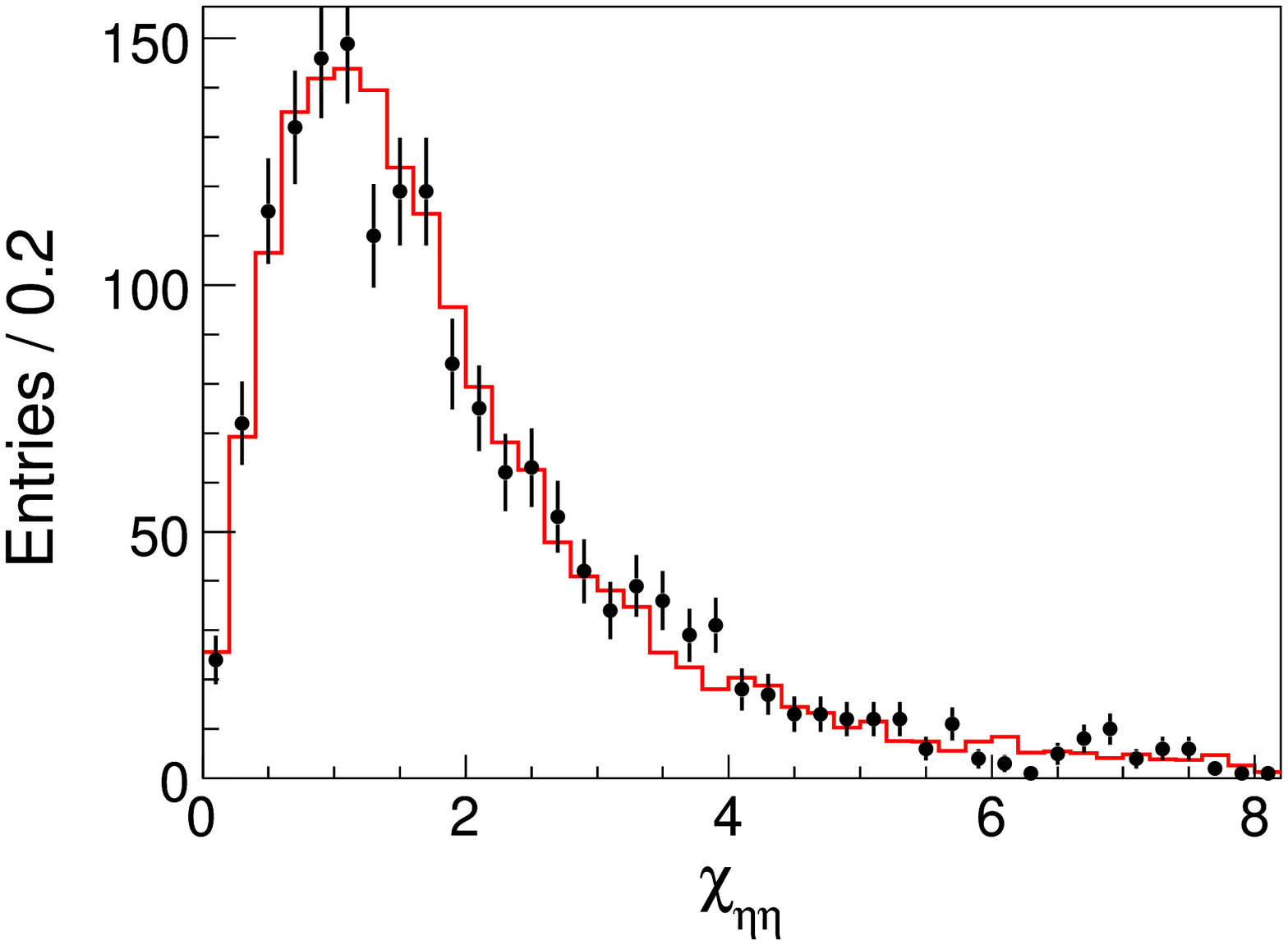,width=8cm,angle=0}
              \put(-30,120){(d)}}
   \caption{Comparisons between data and Monte Carlo simulation for
     $\psi^{\prime}\to\gamma\chi_{cJ}, \chi_{cJ}\to\eta\eta$ for the selection
criteria used.
   (a) Photon multiplicity distribution.
   (b) The $\gamma\gamma$ invariant mass distribution for $\eta$
candidates.
   (c) The distribution of $p_{t\gamma}^2$.
   (d) The $\chi_{\eta\eta}$ distribution.
   Dots with error bars are data in the $\chi_{c0}$
region. The histogram is the Monte Carlo simulation for
$\psi^{\prime} \to \gamma\chi_{c0}, \chi_{c0}\to\eta\eta$ plus the
normalized background estimated from inclusive $\psi^{\prime}$ Monte
Carlo samples.}
   \label{comp_ee}
\end{figure}

Figure~\ref{comp_pp} and Figure~\ref{comp_ee} show comparisons in
the $\chi_{c0}$ region between data and Monte Carlo simulation for
the selection criteria used.  The good agreement across the
distributions shows that the efficiency estimated from Monte Carlo
simulation is reliable.

\section {\boldmath Background analysis}
The backgrounds in the selected event sample from a number of
potential background channels listed in the PDG~\cite{PDG} are
studied with Monte Carlo simulations. The main background to
$\chi_{cJ}\to\pi^0\pi^0$ originates from
$\psi^{\prime}\to\gamma\chi_{cJ}$, $\chi_{cJ}\to\gamma J/\psi$,
$J/\psi\to\gamma\eta$. Using the world average branching
fractions~\cite{PDG} for this mode, we estimate that 48 events from
this channel are in the signal region. However, the simulation also
shows that the background does not peak at the $\chi_{c0}$ nor the
$\chi_{c2}$ mass region. The main backgrounds to
$\chi_{cJ}\to\eta\eta$ originate from $\psi^{\prime}\to\pi^0\pi^0
J/\psi$ and $\psi^{\prime}\to\eta J/\psi$, $J/\psi\to\gamma\eta$.
There are about 233 surviving background events in the signal
region.

\begin{figure}[htb]

   \centerline{
   \psfig{file=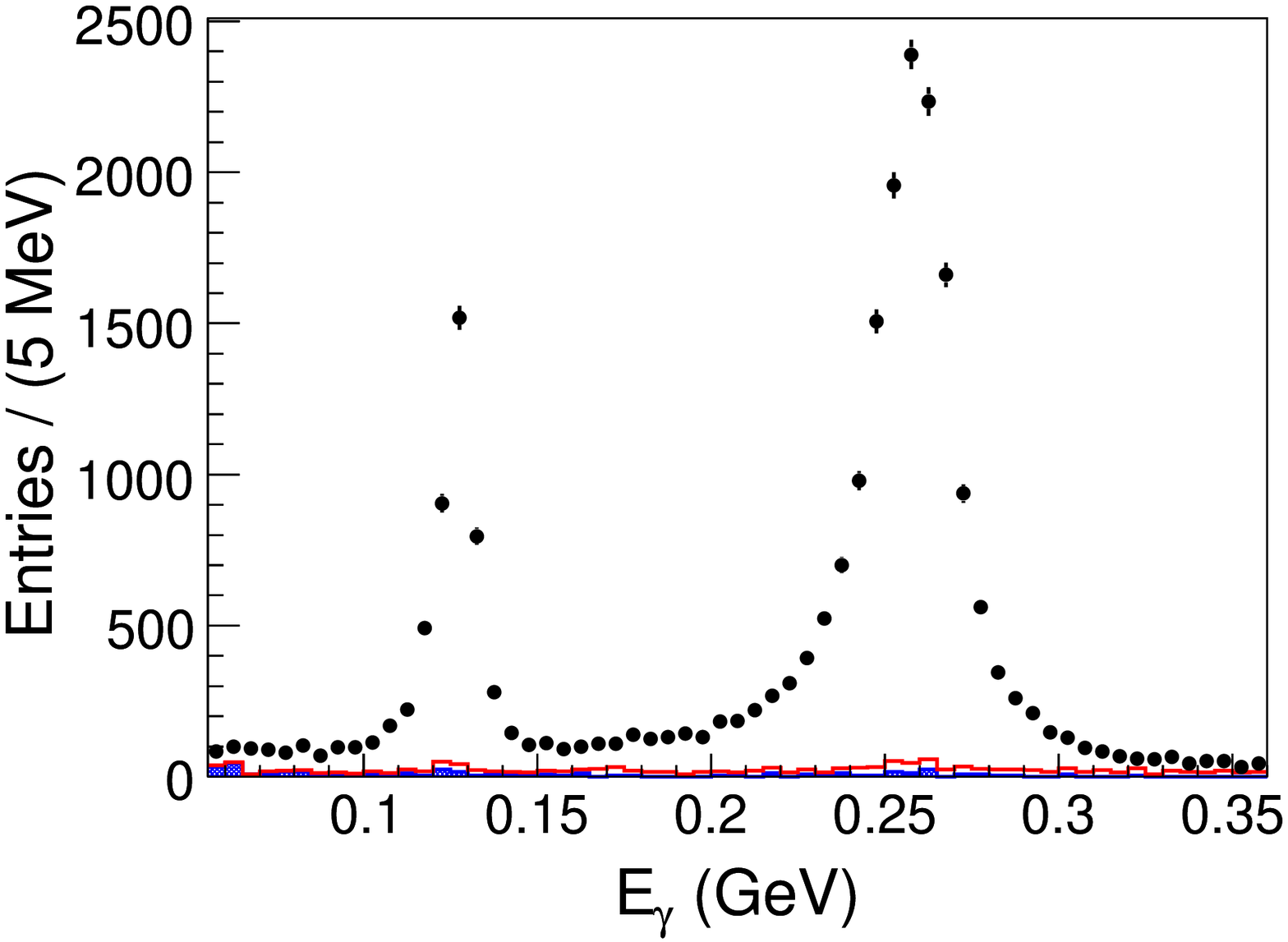,width=8cm,angle=0}
              \put(-30,120){(a)}
   \psfig{file=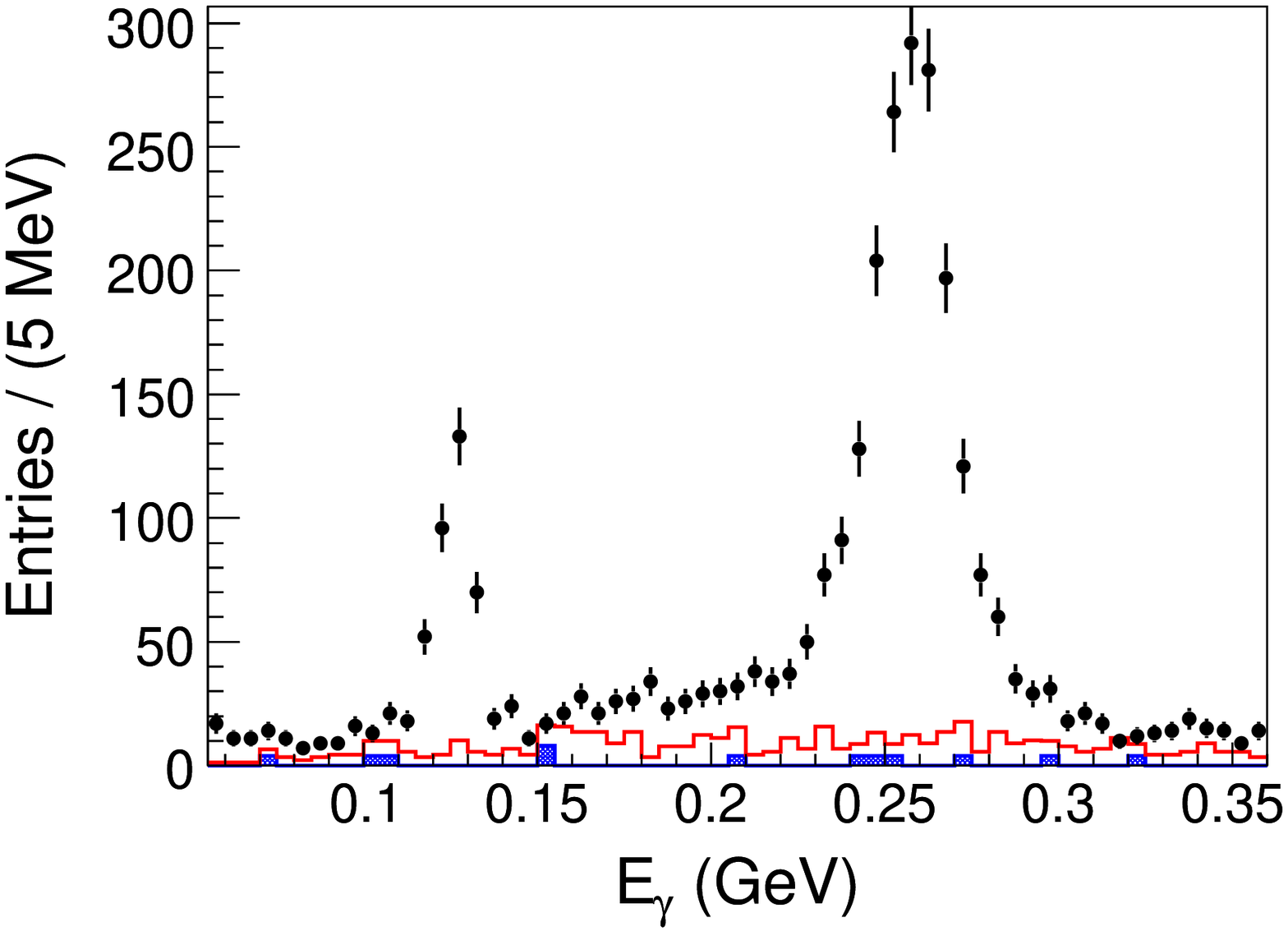,width=8cm,angle=0}
              \put(-30,120){(b)}}
   \caption{
 Radiative photon energy distributions of (a) selected
$\chi_c\to\pi^0\pi^0$ events, and (b) selected $\chi_c\to\eta\eta$ events.
   Dots with error bars are data. The open histogram is the normalized
background estimated from the inclusive $\psi^{\prime}$ Monte Carlo
sample and from the continuum. The shaded histogram is the
normalized contribution from the continuum.}
   \label{incl}
\end{figure}

A $10^8$ inclusive $\psi^{\prime}$ Monte Carlo event sample is also
used to investigate other possible surviving background events.
Figs.~\ref{incl} (a) and \ref{incl} (b) show the radiative photon
energy distribution of the selected $\chi_{cJ}\to\pi^0\pi^0$ and
$\chi_{cJ}\to\eta\eta$ events and the normalized backgrounds
estimated with the inclusive $\psi^{\prime}$ Monte Carlo sample,
respectively.  In the $\chi_{cJ}$ signal region, there is no peaking
background from the inclusive $\psi^{\prime}$ Monte Carlo sample.

The background in our signal region originating from non-resonant
processes is studied using a continuum data sample collected
at a center of mass energy of 3.65~GeV. Normalized
according to the luminosities, the contribution to
$\chi_{cJ}\to\pi^0\pi^0$ is 384 events, as shown in
Fig.~\ref{incl} (a), and the contribution to $\chi_{cJ}\to\eta\eta$ is
48 events, as shown in Fig.~\ref{incl} (b). These backgrounds are small, do not peak in the signal region, and are included as part of the polynomial background below.

\section{\boldmath Number of $\psi^{\prime}$ events}

The number of $\psi^{\prime}$ events, $N_{\psi^{\prime}}$, used in
this analysis is determined from the number of inclusive hadronic
$\psi^{\prime}$ decays. Charged tracks are selected requiring their
point of closest approach to the beam axis be within 1~cm of the
beam line, and their angle with respect to the beam axis, $\theta$,
must satisfy $|\cos \theta| < 0.93$.  Photon candidates must have at
least 25 (50) MeV of energy in the barrel (end-cap) EMC, and have
$|\cos \theta| < 0.93$.

Event selection requires at least one charged track. To remove beam
associated background and background from Bhabha events, there are
special requirements on low charged multiplicity events.  For events
with one charged track, there must be at least three photons, the
acolinearity angle between the two highest energy photons must be
greater than $7^{\circ}$, and the total energy in the EMC, $E_{EMC}$,
must be greater than 0.2 of the center of mass energy, $E_{cm}$, and
less than $0.85 E_{cm}$.  Events with two or three tracks must have
$E_{EMC} > 0.2 E_{cm}$ in order to suppress beam associated
backgrounds. Backgrounds from Bhabha events are reduced by requiring
the presence of at least two photons and $E_{EMC} < 0.85 E_{cm}$ or
the largest energy deposit in the calorimeter less than 0.85 times the
beam energy, $E_{beam}$.  In addition, the second highest momentum
track must have momentum less than $0.9 E_{beam}$, and the
acolinearity angle in the $x-y$ plane of the two highest momentum
tracks must be greater than $7^{\circ}$.

\begin{figure}[htb]
   \centerline{
   \psfig{file=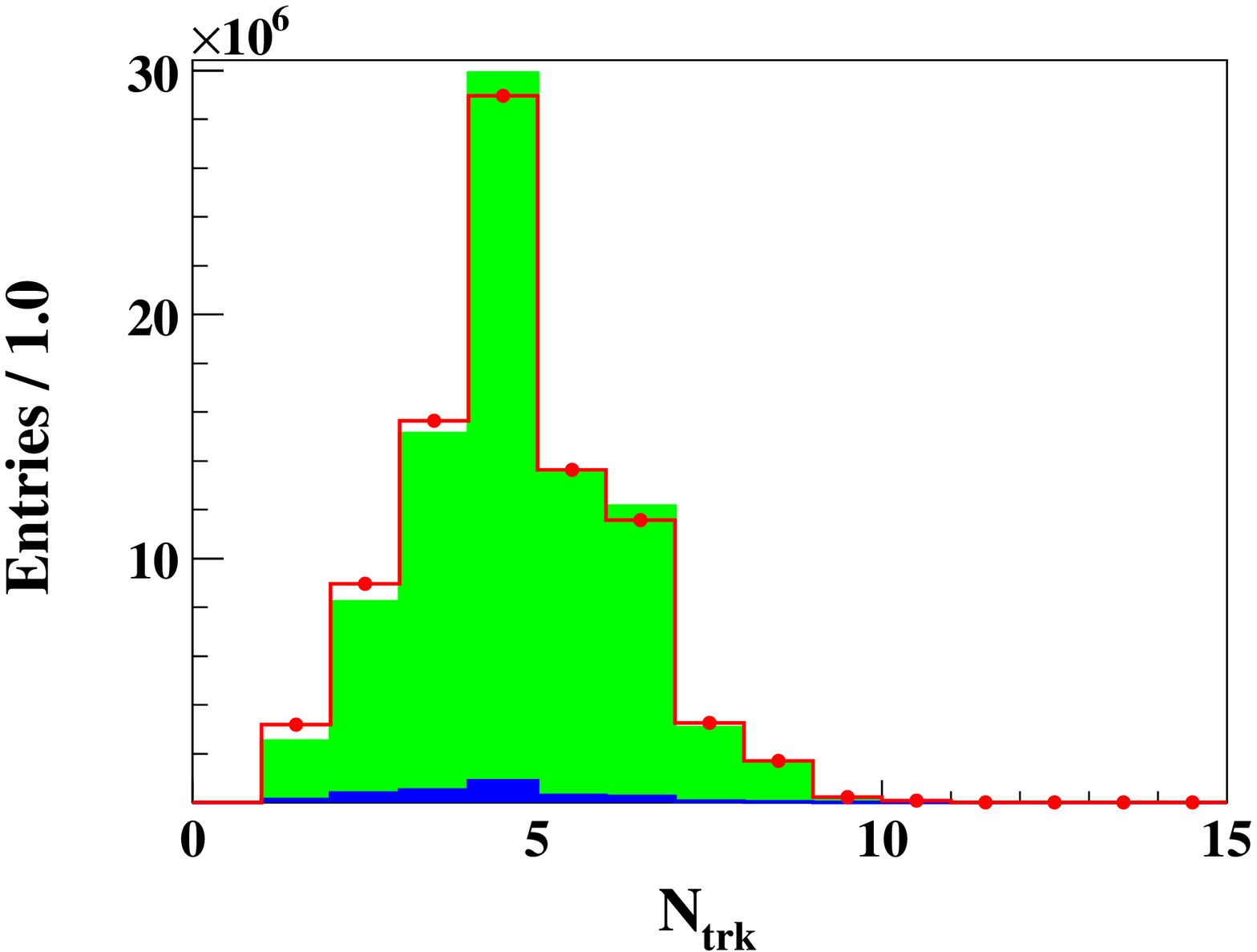,width=8cm,angle=0}
              \put(-30,120){(a)}
   \psfig{file=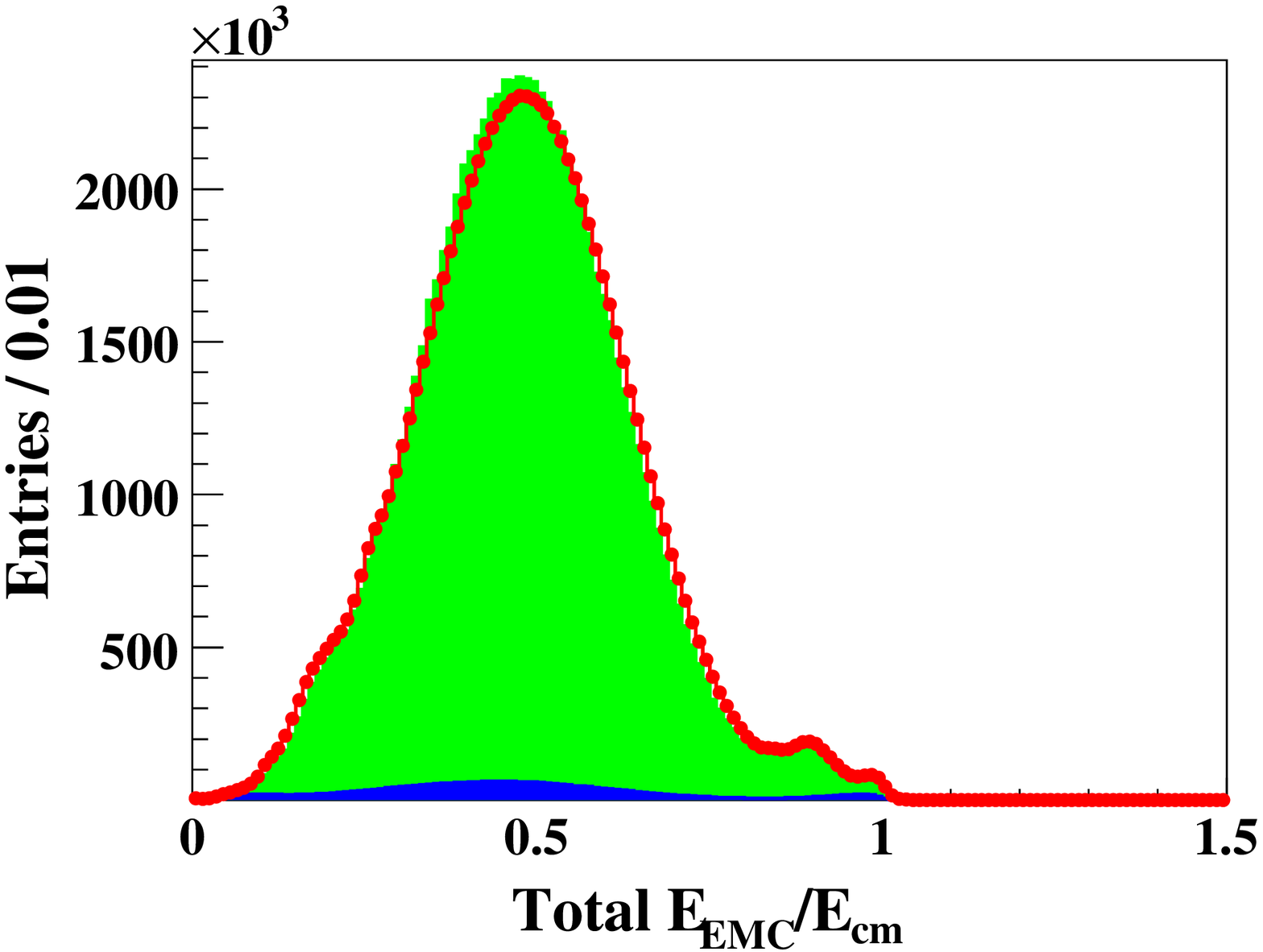,width=8cm,angle=0}
              \put(-30,120){(b)}}

\caption{(a) The distribution of the number of charged tracks for
events satisfying selection criteria. (b) The distribution of the
total energy in the EMC divided by the center of mass energy,
$E_{EMC}/E_{cm}$, for events satisfying selection criteria. All
requirements are applied to events with one to three charged tracks
except the EMC requirements. Dots are data, the light shaded
histogram is the sum of normalized continuum and $\psi^{\prime} \to
hadrons$ simulated events, and the dark shaded histogram is from
continuum data. } \label{psi2snum}
\end{figure}

        The number of hadronic events is determined from the
distribution of $\bar{z}$, which is the average of the distances,
$z$, from the interaction point along the beam of the point of
closest approach of tracks to the beam line. Two methods are used:
fitting the distribution with a Gaussian plus a 2nd order polynomial
background and counting events in a signal region and subtracting
sideband events.  Backgrounds from Bhabha, dimuon, and ditau events
surviving the selection criteria are very small.  The continuum
contribution and the surviving backgrounds are removed by
subtracting the number of events selected with the above criteria
from a continuum sample taken at a center of mass energy of 3.65 GeV
and normalized by relative luminosity and the $1/s$ dependence.  The
efficiency for $\psi^{\prime} \to hadrons$ is determined by
simulation~\cite{kkmc} and is 0.80.  The agreement between data and
Monte Carlo (MC) simulation is shown for the distribution of the
number of charged tracks in Fig.~\ref{psi2snum} (a) and for
$E_{EMC}$ in Fig.~\ref{psi2snum} (b).

The result is $N_{\psi^{\prime}} = (1.06 \pm 0.04)\times 10^8$,
where the error is systematic and is determined mostly by the track
efficiency difference between data and MC (1.2\%), the variation
with the minimum charged track multiplicity requirement (2.86\%),
the difference when a minimum transverse momemtum requirement is
used (0.95\%), the uncertainty of the generator model (0.61\%), and
error due to the continuum subtraction (0.91\%). The statistical
error is negligible. A second analysis using a much different
selection criteria with a higher efficiency determines an almost
identical result.

\section {\boldmath Fitting results}
The $\chi_c\to\pi^0\pi^0$ branching fraction is calculated using
$$Br(\chi_c\to\pi^0\pi^0) =  \frac{N_{obs}}{N_{\psi^{\prime}}\cdot \varepsilon\cdot
Br(\psi^{\prime}\to\gamma\chi_{cJ})\cdot Br(\pi^0\to
\gamma\gamma)\cdot Br(\pi^0\to \gamma\gamma)},$$ where $N_{obs}$ is
the number of events observed, $N_{\psi^{\prime}}$ is the number of
$\psi^{\prime}$ events, and $\varepsilon$ is the selection
efficiency obtained from MC simulation.

\begin{figure}[htb]
   \centerline{
   \psfig{file=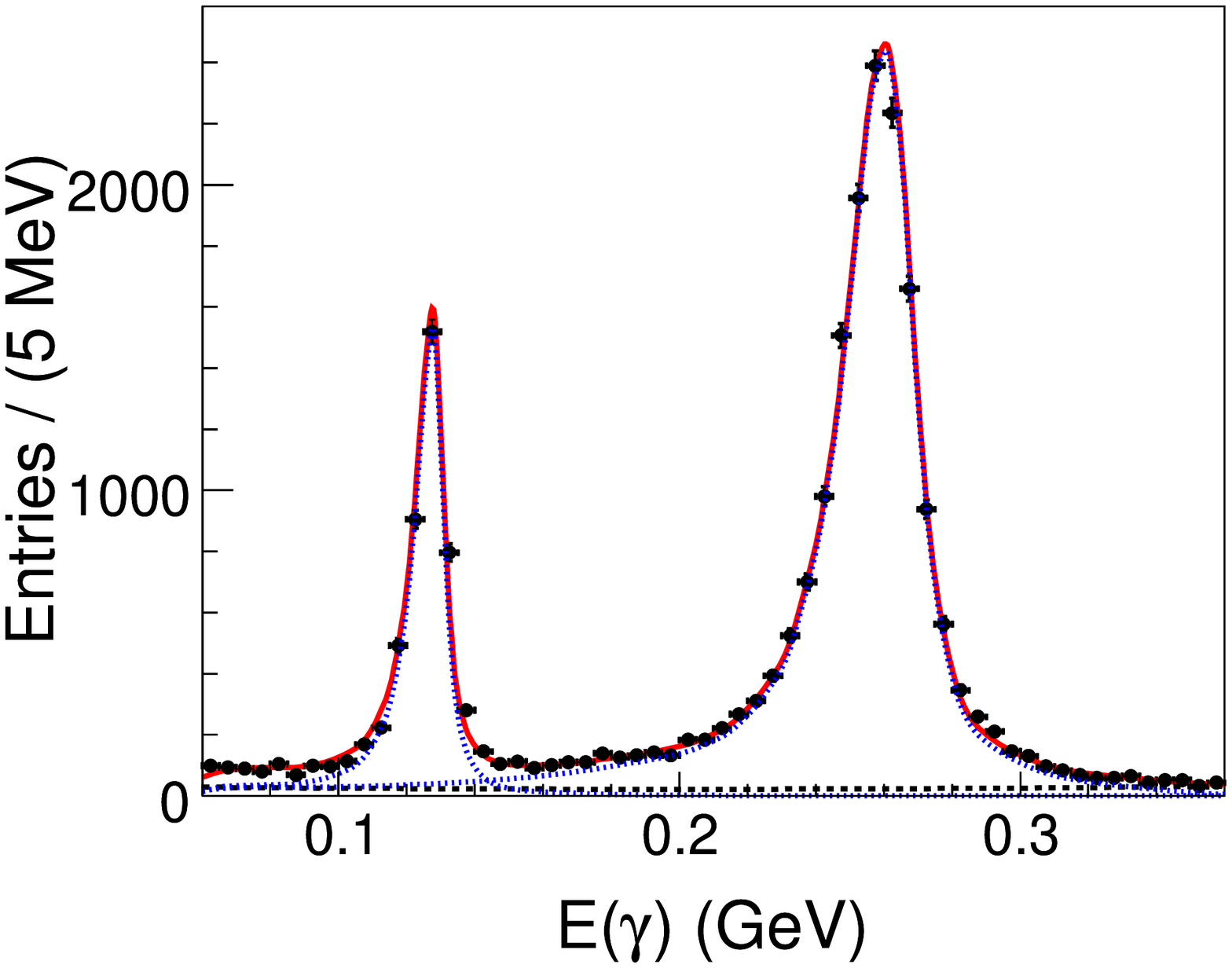,width=8cm,angle=0}
              }

   \caption{The radiative photon energy spectrum of selected
$\chi_c\to\pi^0\pi^0$ events. Dots with error bars are data. The
solid curve is the result of a fit described in the text. The dotted
curves are the $\chi_{cJ}$ signals. The dashed curve is the
background polynomial.}
   \label{gpp_fit}
\end{figure}

The radiative photon energy spectrum of $\chi_{cJ}\to\pi^0\pi^0$
candidates, shown in Fig.~\ref{gpp_fit}, is fitted using an unbinned
maximum likelihood fit in the range from 0.06~GeV to 0.36~GeV. The
shapes of the $\chi_{c0}$ and $\chi_{c2}$ are obtained from
Monte Carlo simulation and the masses and widths of $\chi_{cJ}$ are
fixed to their PDG values~\cite{PDG}. A 2nd-order Chebychev polynomial is used to describe the backgrounds, including those found in the inclusive MC study and the continuum.
The fit gives a $\chi_{c0}$ signal yield of $17443 \pm 167$~events
and a $\chi_{c2}$ signal yield of $4516 \pm 80$~events. The
selection efficiency from Monte Carlo simulation of
$\psi^{\prime}\to
\gamma\chi_{c0}(\chi_{c0}\to\pi^0\pi^0,\pi^0\to\gamma\gamma$) is
$(55.6 \pm 0.2)$\% and the efficiency of $\psi^{\prime}\to
\gamma\chi_{c2}(\chi_{c2}\to\pi^0\pi^0,\pi^0\to\gamma\gamma$) is
$(59.8 \pm 0.2)$\%. The branching fractions are then determined to
be
$$Br(\chi_{c0}\to\pi^0\pi^0)=(3.23\pm0.03)\times 10^{-3},$$
$$Br(\chi_{c2}\to\pi^0\pi^0)=(8.8\pm0.2)\times 10^{-4},$$
where the errors are statistical only.

\begin{figure}[htb]
   \centerline{
   \psfig{file=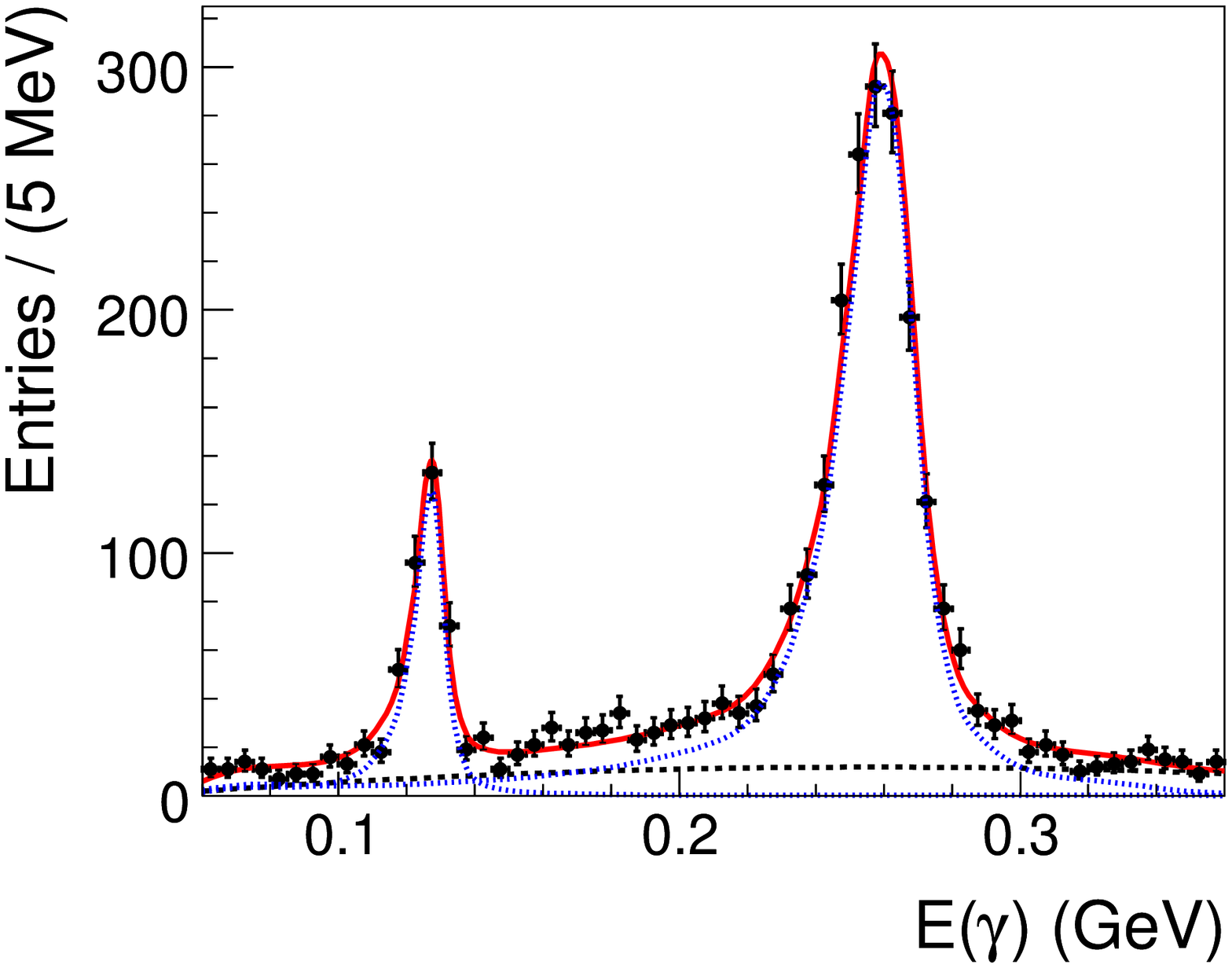,width=8cm,angle=0}
              }

   \caption{The radiative photon energy spectrum of selected
$\chi_c\to\eta\eta$ events. Dots with error bars are data. The solid
curve is the result of a fit described in the text. The dotted
curves are the $\chi_{cJ}$ signals. The dashed curve is the
background polynomial.}
   \label{gee_fit}
\end{figure}

The fit to the radiative photon energy spectrum of
$\chi_{cJ}\to\eta\eta$ candidates, shown in Fig.~\ref{gee_fit}, gives
a $\chi_{c0}$ signal yield of $2132 \pm 60$~events and a $\chi_{c2}$
signal yield of $386 \pm 25$~events. The selection efficiency is $40.3
\pm 0.2$\% and $43.9 \pm 0.2$\% for $\chi_{c0}\to\eta\eta$ and
$\chi_{c2}\to\eta\eta$, respectively. The branching fractions are
$$Br(\chi_{c0}\to\eta\eta)=(3.44\pm0.10)\times 10^{-3},$$
$$Br(\chi_{c2}\to\eta\eta)=(6.5\pm0.4)\times 10^{-4},$$
where the errors are statistical only.
\section {\boldmath Systematic uncertainties}

The systematic uncertainties on the branching fractions come from
many different sources and are summarized in Table~\ref{syserr}. The
uncertainty due to photon detection and photon conversion is 1\% per
photon. This is determined from studies of photon detection
efficiencies in well
understood decays such as $J/\psi\to\rho^0\pi^0$~ and study of photon conversion via $e^+e^-\to\gamma\gamma$~.

The uncertainty due to $\pi^0$ selection is determined from a high
purity control sample of $J/\psi\to\pi^+\pi^-\pi^0$ decays.  The
$\pi^0$ selection efficiency is obtained from the change in the
$\pi^0$ yield in the $\pi^+\pi^-$ recoiling mass spectrum with or
without the $\pi^0$ selection requirement. The difference of $\pi^0$
reconstruction efficiency between data and Monte Carlo simulation
gives an uncertainty of 1\% per $\pi^0$. The uncertainty from the
$\eta$ selection is 1\% per $\eta$, which is determined in a similar
way from a high purity control sample of $J/\psi\to\eta p\bar{p}$.

The systematic error from the $p_{t\gamma}^2$ requirement is
determined by not using the requirement.  The change in the yield
gives systematic errors of 0.9\% for $\chi_{c0}\to\pi^0\pi^0$, 1.2\%
for $\chi_{c2}\to\pi^0\pi^0$, 0.1\% for $\chi_{c0}\to\eta\eta$, and
0.3\% for $\chi_{c2}\to\eta\eta$. The uncertainties from
$\chi_{\eta\eta}$ requirement are 0.6\% for $\chi_{c0}\to\eta\eta$
and 2.6\% for $\chi_{c2}\to\eta\eta$, and are determined in a similar
way.

Since the shapes of the signals in the fit are obtained from Monte Carlo
 simulation, their uncertainties are estimated by changing the masses
 and widths of $\chi_{cJ}$ by one standard
 deviation from the PDG values~\cite{PDG} and taking into account the uncertainties
 of the photon energy scale and resolution in the Monte Carlo simulation.
 They are 1.6\% for
 $\chi_{c0}\to\pi^0\pi^0$, 1.2\% for $\chi_{c2}\to\pi^0\pi^0$, 1.4\%
 for $\chi_{c0}\to\eta\eta$, and 1.5\% $\chi_{c2}\to\eta\eta$.



The background uncertainties are evaluated by changing the background
fitting function from a second order polynomial to third order,
resulting in changes of branching ratios by 0.5\% for
$\chi_{c0}\to\pi^0\pi^0$, 0.5\% for $\chi_{c2}\to\pi^0\pi^0$, 0.2\%
for $\chi_{c0}\to\eta\eta$, and 0.3\% for $\chi_{c2}\to\eta\eta$.

The systematic uncertainties due to the fitting of the radiative
photon energy spectrum were evaluated by changing the fitting range
from (0.05, 0.37)~GeV to (0.07, 0.35)~GeV. The change in yield for
this variation gives systematic uncertainties of 0.3\% for
$\chi_{c0}\to\pi^0\pi^0$, 0.3\% for $\chi_{c2}\to\pi^0\pi^0$, 0.8\% for
$\chi_{c0}\to\eta\eta$, and 1.3\% for $\chi_{c2}\to\eta\eta$.

The systematic uncertainties due to the trigger efficiency in these
neutral channels is estimated to be $<$ 0.1\%, based on cross checks
using different trigger conditions. The uncertainty on the number of
$\psi^{\prime}$ events is 4\%.

The total systematic uncertainties, shown in Table.~\ref{syserr},
are obtained by adding all the above systematic errors in
quadrature. The uncertainty due to the
$\psi^{\prime}\to\gamma\chi_c$ branching fractions is kept separate
and quoted as a second systematic uncertainty.

\begin{table}[htb]
\caption{Systematic uncertainties expressed in percent. }

\begin{tabular}{|l|c|c|c|c|c|}
\hline
Mode &$\chi_{c0}\to\pi^0\pi^0$&$\chi_{c2}\to\pi^0\pi^0$&$\chi_{c0}\to\eta\eta$ &$\chi_{c2}\to\eta\eta$  \\
\hline
photon detection &5 &5 &5 &5 \\
\hline
 $\pi^0$($\eta$) reconstruction  &2 &2 &2 &2 \\
 \hline
  $p_{t\gamma}^2$ & 0.9&1.2 &0.1 &0.3\\
  \hline
   $\chi_{\eta\eta}$ &- &- &0.6 &2.6 \\
   \hline
    signal shape &1.6 &1.2 &1.4 &1.5 \\
    \hline
     background shape & 0.5&0.5 &0.2 &0.3\\
     \hline
      fitting range  &0.3 &0.3 &0.8 &1.3\\
      \hline
       trigger  &0.1 &0.1 & 0.1&0.1\\
       \hline
        $N_{\psi^{\prime}} $&4 & 4& 4&4\\
        \hline
        Total  &7.0 &6.9 &6.9 &7.5 \\
\hline

\end{tabular}
\label{syserr}
\end{table}

\begin{table}[htb]
\caption{Branching fraction results (in units of $10^{-3}$) for each
decay mode. The uncertainties are statistical, systematic due to
this measurement, and systematic due to the branching fractions of
$\psi^{\prime}\to\gamma\chi_{cJ}$, respectively. CLEOc results are
determined using their own branching fractions for $\psi^{\prime}
\to \gamma \chi_{cJ}$, while ours are determined using branching
fractions from the PDG.  If we use the CLEOc branching fractions, we
find $Br(\chi_{c0} \to \pi^0 \pi^0) = 3.29 \times 10^{-3}$,
$Br(\chi_{c0} \to \eta\eta) = 3.51 \times 10^{-3}$, $Br(\chi_{c2}
\to \pi^0 \pi^0) = 0.78 \times 10^{-3}$, and $Br(\chi_{c2} \to
\eta\eta) = 0.58 \times 10^{-3}$.}

\begin{tabular}{|l|c|c|c|}

\hline
Mode  &  & $\chi_{c0}$ &  $\chi_{c2}$ \\
\hline

$\pi^0\pi^0$
 & $\ $This Work$\ $ & $3.23\pm0.03\pm0.23\pm0.14$ & $0.88\pm0.02\pm0.06\pm0.04$ \\
 & $\ $CLEOc~\cite{bib:cleoc}$\ $ & $2.94\pm0.07\pm0.32\pm0.15$ & $0.68\pm0.03\pm0.07\pm0.04$ \\
 & PDG~\cite{PDG}    & $2.43\pm 0.20$             &  $0.71\pm0.08$  \\

$\eta\eta$
 &$\ $ This Work$\ $ & $3.44\pm0.10\pm0.24\pm0.13$ & $0.65\pm0.04\pm0.05\pm0.03$ \\
 &$\ $ CLEOc~\cite{bib:cleoc}$\ $ & $3.18\pm0.13\pm0.31\pm0.16$ & $0.51\pm0.05\pm0.05\pm0.03$ \\
 & PDG~\cite{PDG}        &    $2.4\pm0.4$ & $<0.5$  \\

\hline
\end{tabular}
\label{br}
\end{table}
\section {\boldmath Summary}
In summary, with a sample of $1.06 \times 10^8~\psi^{\prime}$ events
in the BESIII detector, improved measurements of the branching
fractions of $\chi_{c0,2}\to\pi^0\pi^0$ and $\chi_{c0,2}\to\eta\eta$
are performed: ~$Br(\chi_{c0}\to\pi^0\pi^0)=(3.23\pm 0.03\pm0.23 \pm
0.14)\times 10^{-3}$,~$Br(\chi_{c2}\to\pi^0\pi^0)=(8.8\pm 0.2\pm
0.6\pm0.4 )\times 10^{-4}$,~$Br(\chi_{c0}\to\eta\eta)=(3.44\pm
0.10\pm0.24 \pm0.2 )\times 10^{-3}$, and
$Br(\chi_{c2}\to\eta\eta)=(6.5\pm 0.4\pm 0.5\pm 0.3)\times 10^{-4}$,
where the uncertainties are statistical, systematic due to this
measurement, and systematic due to the branching fractions of
$\psi^{\prime}\to\gamma\chi_{cJ}$, respectively. Results are listed
in Table~\ref{br} and compared with previous measurements.

\section{ \boldmath Acknowledgments}
The BESIII collaboration thanks the staff of BEPC and the computing
center for their hard efforts. This work is supported in part by the
Ministry of Science and Technology of China under Contract No.
2009CB825200; National Natural Science Foundation of China (NSFC)
under Contracts Nos. 10625524, 10821063, 10825524, 10835001,
10935007; the Chinese Academy of Sciences (CAS) Large-Scale
Scientific Facility Program; CAS under Contracts Nos. KJCX2-YW-N29,
KJCX2-YW-N45; 100 Talents Program of CAS; Istituto Nazionale di
Fisica Nucleare, Italy; Russian Foundation for Basic Research under
Contracts Nos. 08-02-92221, 08-02-92200-NSFC-a; Siberian Branch of
Russian Academy of Science, joint project No 32 with CAS; the
Chinese University of Hong Kong Focused Investment Grant under
Contract No. 3110031; U. S. Department of Energy under Contracts
Nos. DE-FG02-04ER41291, DE-FG02-91ER40682, DE-FG02-94ER40823; WCU
Program of National Research Foundation of Korea under Contract No.
R32-2008-000-10155-0. D. Cronin-Hennessy thanks the A.P. Sloan
Foundation. This paper is also supported by the NSFC under Contract
Nos. 10979038, 10875113.

\end{document}